\documentclass[onecolumn,preprintnumbers,pra,hyperref]{revtex4}
\usepackage{amssymb}
\usepackage{amsfonts}
\usepackage{amsmath}
\usepackage{graphicx}

\begin{document}

\title{New identities about operator Hermite polynomials and their related
integration formulas\thanks{Work supported by the National Natural
Science Foundation of China under grant: 10775097 and 10947017/A05,
and by the specialized research fund for the doctorial progress of
higher education of China (No: 20070358009) }}
\author{Hong-Yi Fan and Hong-Chun Yuan$^{*}$
\\$^{1}${\small Department of Physics, Shanghai Jiao Tong University, Shanghai,
200030, China}\\$^{*}${\small Corresponding author. E-mail:
yuanhch@126.com or yuanhch@sjtu.edu.cn}}

\begin{abstract}
By virtue of the technique of integration within an ordered product (IWOP) of
operators and the bipartite entangled state representation we derive some new
identities about operator Hermite polynomials in both single- and
two-variable, we also find a binomial-like theorem between the single-variable
Hermite polynomials and the two-variable Hermite polynomials. Application of
these identities in deriving new integration formulas, but without really
doing the integration in the usual sense, is demonstrated.

\textbf{Keywords:} IWOP technique; operator Hermite polynomials; integration formulas

\end{abstract}
\maketitle

Hermite polynomials $H_{n}\left(  x\right)  $ are frequently used in
quantum mechanics and mathematical physics\cite{r1,r2,r3}, for
instance, the wave function of number state (Fock state) is
$\psi_{n}\left(  x\right)  =\left( 2^{n}n!\sqrt{\pi}\right)
^{-1/2}e^{-x^{2}/2}H_{n}\left(  x\right)  $. Here we shall focus on
operator Hermite polynomials, because they are very useful in
miscellaneous calculations in quantum mechanics. Due to the fact
that operators in quantum mechanics can be expressed by Dirac's
ket-bra $\left \vert \left.  {}\right.  \right \rangle \left \langle
\left.  {}\right.  \right \vert $, so the research of operator
Hermite polynomials should be closely related to quantum mechanical
representations. In Refs.\cite{r4,r5,r6}, we have
derived the following identities about operator Hermite polynomials%
\begin{equation}
H_{n}\left(  X\right)  =2^{n}\colon X^{n}\colon \label{1}%
\end{equation}
and%
\begin{equation}
X^{n}=\left(  2i\right)  ^{-n}\colon H_{n}\left(  \text{i}X\right)
\colon,\label{2}%
\end{equation}
where $\colon \colon$ denotes normal ordering, and $X=(a+a^{\dag})/\sqrt{2}$ is
the coordinate operator with $\left[  a,a^{\dag}\right]  =1$. These are very
useful identities. For example, $H_{n}\left(  X\right)  $ operating on the
vacuum state $\left \vert 0\right \rangle $ yields
\begin{equation}
H_{n}\left(  X\right)  \left \vert 0\right \rangle =2^{n/2}a^{\dagger
n}\left \vert 0\right \rangle =\sqrt{n!2^{n}}\left \vert n\right \rangle ,
\end{equation}
where $\left \vert n\right \rangle =a^{\dagger n}/\sqrt{n!}\left \vert
0\right \rangle $ is the number state, then using the coordinate eigenvector
equation $X\left \vert x\right \rangle =x\left \vert x\right \rangle ,$ we
immediately have the relation between $\left \langle x\right.  \left \vert
n\right \rangle $ and $\left \langle x\right \vert \left.  0\right \rangle ,$
\begin{equation}
\left \langle x\right.  \left \vert n\right \rangle =\frac{1}{\sqrt{n!2^{n}}%
}\left \langle x\right \vert H_{n}\left(  X\right)  \left \vert 0\right \rangle
=\frac{1}{\sqrt{n!2^{n}}}H_{n}\left(  x\right)  \left \langle x\right \vert
\left.  0\right \rangle .\text{ }\label{3}%
\end{equation}
Eq.(\ref{1}) can provide some new integration formulas in a direct manner, for
instance, from
\begin{align}
\int H_{n}\left(  X\right)  dX &  =2^{n}\int \colon X^{n}\colon dX=\frac{2^{n}%
}{n+1}\colon X^{n+1}\colon+C\nonumber \\
&  =\frac{1}{2\left(  n+1\right)  }H_{n+1}\left(  X\right)  +C,\label{4}%
\end{align}
we easily see%
\begin{equation}
\int_{0}^{y}H_{n}\left(  x\right)  dx=\frac{1}{2\left(  n+1\right)  }\left[
H_{n+1}\left(  y\right)  -H_{n+1}\left(  0\right)  \right]  .\label{5}%
\end{equation}

Further, using the technique of integration within an ordered product (IWOP)
of operators\cite{r7,r8,r9} and the completeness relation of $\left \vert
x\right \rangle $, $\int dx\left \vert x\right \rangle \left \langle x\right \vert
=\int \frac{dx}{\sqrt{\pi}}\colon e^{-\left(  x-\hat{X}\right)  ^{2}}\colon=1$,
it follows from Eq.(\ref{1}) that%
\begin{equation}
H_{n}\left(  \hat{X}\right)  =\int_{-\infty}^{\infty}\frac{dx}{\sqrt{\pi}%
}\colon e^{-\left(  x-X\right)  ^{2}}\colon H_{n}\left(  x\right)
=2^{n}\colon X^{n}\colon \label{6}%
\end{equation}
and%
\begin{equation}
X^{n}=\int \frac{dx}{\sqrt{\pi}}x^{n}\colon e^{-\left(  x-X\right)  ^{2}}%
\colon=\left(  2i\right)  ^{-n}\colon H_{n}\left(  \text{i}X\right)
\colon,\label{7}%
\end{equation}
which respectively implies the integration formulas\cite{r10} (we have the
result without really performing the integration in the usual sense)
\begin{equation}
\int \frac{dx}{\sqrt{\pi}}e^{-\left(  x-y\right)  ^{2}}H_{n}\left(  x\right)
=2^{n}y^{n}\label{8}%
\end{equation}
and%
\begin{equation}
\int \frac{dx}{\sqrt{\pi}}e^{-\left(  x-y\right)  ^{2}}x^{n}=\left(  2i\right)
^{-n}H_{n}\left(  \text{i}y\right)  .\label{9}%
\end{equation}
Similarly, for the momentum representation's completeness relation%
\begin{equation}
\int dp\left \vert p\right \rangle \left \langle p\right \vert =\int \frac
{dp}{\sqrt{\pi}}\colon e^{-\left(  p-P\right)  ^{2}}\colon=1
\end{equation}
where $P\left \vert p\right \rangle =p\left \vert p\right \rangle $, and
$P=\frac{a-a^{\dagger}}{i\sqrt{2}},$ we have%
\begin{equation}
H_{n}\left(  P\right)  =\int \frac{dp}{\sqrt{\pi}}\colon e^{-\left(
p-P\right)  ^{2}}\colon H_{n}\left(  p\right)  =2^{n}\colon P^{n}\colon.
\end{equation}
It then follows%
\begin{equation}
H_{n}\left(  P\right)  \left \vert 0\right \rangle =i^{n}\sqrt{n!2^{n}%
}\left \vert n\right \rangle
\end{equation}
and%
\begin{equation}
\left \langle p\right.  \left \vert n\right \rangle =\frac{1}{\sqrt{n!2^{n}}%
}\left \langle p\right \vert H_{n}\left(  P\right)  \left \vert 0\right \rangle
=\frac{\left(  -i\right)  ^{n}}{\sqrt{n!2^{n}}}H_{m}\left(  p\right)
\left \langle p\right \vert \left.  0\right \rangle .
\end{equation}
The above examples shows that re-ordering operator Hermite
polynomials (including two-variable Hermite polynomials) together
with the IWOP technique may work in concisely deriving some new
operator identities and new integration formulas. In the following
we shall proceed in this direction, and we also develop this method
with the use of the entangled state representation.

To begin with, let us firstly point out a misleading, i.e., one
might think that since $H_{n}\left(  X\right)  =2^{n}\colon
X^{n}\colon,$ then $H_{n}\left(  fX\right)  =2^{n}\colon \left(
fX\right)  ^{n}\colon,$ but this
is wrong, because $\left[  a,a^{\dagger}\right]  =1$ in $X=(a+a^{\dag}%
)/\sqrt{2}$, while $a$ and $a^{\dagger}$ are commutable in $\colon X^{n}%
\colon$. In fact, from the generating function formula of\ Hermite
polynomials
\begin{equation}
e^{-t^{2}+2tx}=\sum_{n=0}^{\infty}\frac{t^{n}}{n!}H_{n}\left(  x\right)
,\label{a2}%
\end{equation}
and the Baker-Hausdorff formula%
\begin{equation}
e^{A}e^{B}=e^{A+B}e^{\frac{1}{2}\left[  A,B\right]  },\text{ \ for }\left[
A,\left[  A,B\right]  \right]  =\left[  B,\left[  A,B\right]  \right]
=0,\label{10}%
\end{equation}
as well as the property that Bose operators are permuted within $:$ $:$, for
$f\neq1$ we have%

\begin{align}
\sum_{n=0}^{\infty}\frac{t^{n}}{n!}H_{n}\left(  fX\right)   &  =e^{-t^{2}%
+2tfX}=\colon e^{-\left(  t\sqrt{1-f^{2}}\right)  ^{2}+2\left(  t\sqrt
{1-f^{2}}\right)  \frac{fX}{\sqrt{1-f^{2}}}}\colon \nonumber \\
&  =\sum_{n=0}^{\infty}\frac{\left(  t\sqrt{1-f^{2}}\right)  ^{n}}{n!}\colon
H_{n}\left(  \frac{fX}{\sqrt{1-f^{2}}}\right)  \colon.\label{c9}%
\end{align}
Comparing the coefficients of $t^{n}$ on the two sides we obtain the following
identity%
\begin{equation}
H_{n}\left(  fX\right)  =\left(  \sqrt{1-f^{2}}\right)  ^{n}\colon
H_{n}\left(  \frac{fX}{\sqrt{1-f^{2}}}\right)  \colon \neq2^{n}\colon \left(
fX\right)  ^{n}\colon.\label{c10}%
\end{equation}
Thus we should be very cautious to tackle operator Hermite polynomials. Based
on Eq.(\ref{c10}) and using the coordinate representation's completeness
relation as well as the IWOP technique we have%
\begin{equation}
H_{n}\left(  fX\right)  =\int_{-\infty}^{\infty}\frac{dx}{\sqrt{\pi}}%
H_{n}\left(  fx\right)  \colon e^{-\left(  x-\hat{X}\right)  ^{2}}%
\colon=\left(  1-f^{2}\right)  ^{n/2}\colon H_{n}\left(  \frac{fX}%
{\sqrt{1-f^{2}}}\right)  \colon,\label{c11}%
\end{equation}
which implies the integration formula\cite{r10}%
\begin{equation}
\int_{-\infty}^{\infty}\frac{dx}{\sqrt{\pi}}H_{n}\left(  fx\right)
e^{-\left(  x-y\right)  ^{2}}=\left(  1-f^{2}\right)  ^{n/2}H_{n}\left(
\frac{fy}{\sqrt{1-f^{2}}}\right)  ,\label{c12}%
\end{equation}
so we obtain it without really doing the integration in the usual sense.

Now we turn to the operator Hermite polynomials $H_{n}\left(  \frac{X+Y}%
{\sqrt{2}}\right)  ,$ where $Y=\frac{b+b^{\dag}}{\sqrt{2}}$ is another
coordinate operator with $\left[  b,b^{\dag}\right]  =1,$ and $\left[
X,Y\right]  =0$, we want to derive its normally ordered expansion. Using
Eqs.(\ref{a2}) and (\ref{10}), we also have%
\begin{equation}
\colon e^{2t(X+Y)}\colon=e^{2t(X+Y)－ 2t^{2}}=\sum_{n=0}^{\infty}%
\frac{\left(  t\sqrt{2}\right)  ^{n}}{n!}H_{n}\left(  \frac{X+Y}{\sqrt{2}%
}\right)  .\label{a1}%
\end{equation}
On the other hand,
\begin{equation}
\colon e^{2t(X+Y)}\colon=\sum_{n=0}^{\infty}\frac{\left(  2t\right)  ^{n}}%
{n!}\left.  \colon \left(  X+Y\right)  ^{n}\colon \right.  .\label{a3}%
\end{equation}
Comparing the $t^{n}$ terms in Eq.(\ref{a1}) and Eq.(\ref{a3}) we obtain a new
operator Hermite polynomials identity%
\begin{equation}
H_{n}\left(  \frac{X+Y}{\sqrt{2}}\right)  =2^{n/2}\colon \left(  X+Y\right)
^{n}\colon,\label{a4}%
\end{equation}
which is regarded as the extension of Eq.(\ref{1}). This can also be derived
by virtue of the entangled state representation. Recall that the bipartite
entanglement state representation $\left \vert \xi \right \rangle $ is expressed
as\cite{r11}
\begin{equation}
\left \vert \xi \right \rangle =e^{-\frac{1}{2}\left \vert \xi \right \vert ^{2}+\xi
a^{\dagger}+\xi^{\ast}b^{\dagger}-a^{\dagger}b^{\dagger}}\left \vert
00\right \rangle ,\text{ }\xi=\xi_{1}+i\xi_{2},\label{c1}%
\end{equation}
where $\left \vert 00\right \rangle $ is the two-mode vacuum state, obeys the
eigenvector equations%
\begin{equation}
\left(  a+b^{\dagger}\right)  \left \vert \xi \right \rangle =\xi \left \vert
\xi \right \rangle ,\text{ }\left(  a^{\dag}+b\right)  \left \vert \xi
\right \rangle =\xi^{\ast}\left \vert \xi \right \rangle ,\label{c2}%
\end{equation}
or
\begin{equation}
\left(  X+Y\right)  \left \vert \xi \right \rangle =\sqrt{2}\xi_{1}\left \vert
\xi \right \rangle ,\text{ }\left(  P_{x}-P_{y}\right)  \left \vert
\xi \right \rangle =\sqrt{2}\xi_{2}\left \vert \xi \right \rangle .\label{c3}%
\end{equation}
where $P_{x}=\frac{1}{i\sqrt{2}}(a-a^{\dag})$, and $P_{y}=\frac{1}{i\sqrt{2}%
}(b-b^{\dag}).$ Using $\left \vert 00\right \rangle \left \langle 00\right \vert
=\colon \exp \left(  -a^{\dagger}a-b^{\dag}b\right)  \colon$ and the IWOP
technique we can prove that $\left \vert \xi \right \rangle $ is orthonormal and
complete,
\begin{align}
\int \frac{d^{2}\xi}{\pi}\left \vert \xi \right \rangle \left \langle
\xi \right \vert  &  =\int \frac{d^{2}\xi}{\pi}\colon e^{-\left(  a^{\dagger
}+b-\xi^{\ast}\right)  \left(  a+b^{\dag}-\xi \right)  }\colon \nonumber \\
&  =\int \frac{d\xi_{1}d\xi_{2}}{\pi}\colon e^{-\left(  \xi_{1}-\frac
{X+Y}{\sqrt{2}}\right)  ^{2}-\left(  \xi_{2}-\frac{P_{x}-P_{y}}{\sqrt{2}%
}\right)  ^{2}}\colon=1.\label{c4}%
\end{align}
Based on Eqs.(\ref{c3}) and (\ref{c4}), using the integration formulas
Eq.(\ref{8}), we obtain%
\begin{align}
H_{n}\left(  \frac{X+Y}{\sqrt{2}}\right)   &  =\int \frac{d^{2}\xi}{\pi}%
H_{n}\left(  \xi_{1}\right)  \left \vert \xi \right \rangle \left \langle
\xi \right \vert \nonumber \\
&  =\int \frac{d\xi_{1}d\xi_{2}}{\pi}H_{n}\left(  \xi_{1}\right)  \colon
e^{-\left(  \xi_{1}-\frac{X+Y}{\sqrt{2}}\right)  ^{2}-\left(  \xi_{2}%
-\frac{P_{x}-P_{y}}{\sqrt{2}}\right)  ^{2}}\colon \nonumber \\
&  =\int \frac{d\xi_{1}}{\sqrt{\pi}}H_{n}\left(  \xi_{1}\right)  \colon
e^{-\left(  \xi_{1}-\frac{X+Y}{\sqrt{2}}\right)  ^{2}}\colon \nonumber \\
&  =\left(  \sqrt{2}\right)  ^{n}\colon \left(  X+Y\right)  ^{n}\colon
.\label{c5}%
\end{align}
According to Eq.(\ref{a4}), considering the two-mode coordinate eigenvector'
completeness relation
\begin{equation}
\int dxdy\left \vert x,y\right \rangle \left \langle x,y\right \vert =\int
\frac{dxdy}{\pi}\colon e^{-(x-X)^{2}-(y-Y)^{2}}\colon=1\label{c6}%
\end{equation}
we have%
\begin{align}
H_{n}\left(  \frac{\hat{X}+\hat{Y}}{\sqrt{2}}\right)   &  =\int \frac{dxdy}%
{\pi}H_{n}\left(  \frac{x+y}{\sqrt{2}}\right)  \colon e^{-(x-\hat{X}%
)^{2}-(y-\hat{Y})^{2}}\colon \nonumber \\
&  =\sqrt{2^{n}}\colon \left(  \hat{X}+\hat{Y}\right)  ^{n}\colon,\label{c7}%
\end{align}
which indicates the following new integration formula%

\begin{equation}
\int \frac{dxdy}{\pi}H_{n}\left(  \frac{x+y}{\sqrt{2}}\right)  e^{-(x-\mu
)^{2}-(y-\nu)^{2}}=\left(  \sqrt{2}\mu+\sqrt{2}\nu \right)  ^{n}. \label{c8}%
\end{equation}

Moreover, with the aid of the following sum of the Hermite
polynomials\cite{r10}%
\begin{equation}
\sum_{n=0}^{m}\left(
\begin{array}
[c]{c}%
m\\
n
\end{array}
\right)  H_{m-n}(\sqrt{2}fx)H_{n}(\sqrt{2}gy)=2^{m/2}H_{m}(fx+gy),\label{d1}%
\end{equation}
and using Eqs.(\ref{c12}) and (\ref{c6}), we obtain%
\begin{align}
H_{m}(fX+gY) &  =\int dxdyH_{m}(fx+gy)\left \vert x,y\right \rangle \left \langle
x,y\right \vert \nonumber \\
&  =\int \frac{dxdy}{\pi}H_{m}(fx+gy)\colon e^{-(x-X)^{2}-(y-Y)^{2}}%
\colon \nonumber \\
&  =2^{-m/2}\sum_{n=0}^{m}\left(
\begin{array}
[c]{c}%
m\\
n
\end{array}
\right)  \int \frac{dxdy}{\pi}H_{m-n}(\sqrt{2}fx)H_{n}(\sqrt{2}gy)\colon
e^{-(x-\hat{X})^{2}-(y-\hat{Y})^{2}}\colon \nonumber \\
&  =2^{-m/2}\sum_{n=0}^{m}\left(
\begin{array}
[c]{c}%
m\\
n
\end{array}
\right)  \left(  1-2f^{2}\right)  ^{\left(  m-n\right)  /2}\left(
1-2g^{2}\right)  ^{n/2}\nonumber \\
&  \times \colon H_{m-n}\left(  \frac{\sqrt{2}fX}{\sqrt{1-2f^{2}}}\right)
H_{n}\left(  \frac{\sqrt{2}gY}{\sqrt{1-2g^{2}}}\right)  \colon,\label{d2}%
\end{align}
then operating the sum $\sum_{m=0}^{\infty}\frac{t^{m}}{m!}$ on both sides of
Eq.(\ref{d2}) and using Eq.(\ref{a2}) we obtain%
\begin{align}
&  \sum_{m=0}^{\infty}\frac{t^{m}}{m!}H_{m}(fX+gY)\nonumber \\
&  =\sum_{k=0}^{\infty}\sum_{n=0}^{\infty}\frac{\left(  t\sqrt{\frac{1-2f^{2}%
}{2}}\right)  ^{k}\left(  t\sqrt{\frac{1-2g^{2}}{2}}\right)  ^{n}}{n!k!}\\
&  \times \colon H_{k}\left(  \frac{\sqrt{2}fX}{\sqrt{1-2f^{2}}}\right)
H_{n}\left(  \frac{\sqrt{2}gY}{\sqrt{1-2g^{2}}}\right)  \colon \nonumber \\
&  =\colon \exp \left[  -t^{2}\left(  1-f^{2}-g^{2}\right)  +2t\left(
fX+gY\right)  \right]  \colon \nonumber \\
&  =\sum_{m=0}^{\infty}\frac{\left(  t\sqrt{1-f^{2}-g^{2}}\right)  ^{m}}%
{m!}\colon H_{m}\left(  \frac{fX+gY}{\sqrt{1-f^{2}-g^{2}}}\right)
\colon,\label{d3}%
\end{align}
where we have used the summation formula
\begin{equation}
\sum_{m=0}^{\infty}\sum_{n=0}^{m}A_{m-n}B_{n}=\sum_{k=0}^{\infty}\sum
_{n=0}^{\infty}A_{k}B_{n}.\label{d4}%
\end{equation}
Comparing the $t^{n}$ terms on both sides of Eq.(\ref{d3}), we obtain another
new operator Hermite polynomials identity%
\begin{equation}
H_{m}(fX+gY)=\left(  \sqrt{1-f^{2}-g^{2}}\right)  ^{m}\colon H_{m}\left(
\frac{fX+gY}{\sqrt{1-f^{2}-g^{2}}}\right)  \colon,\label{d5}%
\end{equation}
which is entirely different from Eq.(\ref{a4}), noting that the convergent
condition of Eq.(\ref{d5}) is ,$f=g\neq \frac{1}{\sqrt{2}}.$In particular, when
$f=g=1$, Eq.(\ref{d5}) becomes%
\begin{equation}
H_{m}(X+Y)=i^{m}\colon H_{m}\left[  -i\left(  X+Y\right)  \right]
\colon.\label{d6}%
\end{equation}
which is also a new operator Hermite polynomials. Similarly, according to
Eq.(\ref{c6}), we have
\begin{align}
H_{m}(fX+gY) &  =\int \frac{dxdy}{\pi}H_{m}(fx+gy)\colon e^{-(x-X)^{2}%
-(y-Y)^{2}}\colon \nonumber \\
&  =\left(  \sqrt{1-f^{2}-g^{2}}\right)  ^{m}\colon H_{m}\left(  \frac
{fX+gY}{\sqrt{1-f^{2}-g^{2}}}\right)  \colon,\label{d7}%
\end{align}
this leads to the new integration formula
\begin{equation}
\int \frac{dxdy}{\pi}H_{m}(fx+gy)e^{-(x-\mu)^{2}-(y-\nu)^{2}}=\left(
\sqrt{1-f^{2}-g^{2}}\right)  ^{m}H_{m}\left(  \frac{f\mu+g\nu}{\sqrt
{1-f^{2}-g^{2}}}\right)  .\label{d8}%
\end{equation}
As the special case of Eq.(\ref{d8}), when $f=g=1,$ we see
\begin{equation}
\int \frac{dxdy}{\pi}H_{m}(x+y)e^{-(x-\mu)^{2}-(y-\nu)^{2}}=i^{m}H_{m}\left[
-i\left(  \mu+\nu \right)  \right]  .\label{d9}%
\end{equation}
As can be seen from the above discussion, some integral formulas are
derived without actually performing the integration in the usual
sense, which is the advantage of the IWOP technique.

Finally, according to the generating function of two-variable Hermite
polynomials $H_{m,n}\left(  \xi,\xi^{\ast}\right)  $%
\begin{equation}
\sum_{m,n=0}^{\infty}\frac{t^{m}t^{\prime n}}{m!n!}H_{m,n}\left(  \xi
,\xi^{\ast}\right)  =e^{-tt^{\prime}+t\xi+t^{\prime}\xi^{\ast}},\label{d10}%
\end{equation}
by noticing $\left[  a+b^{\dagger},a^{\dag}+b\right]  =0,$ we have
\begin{align}
e^{-tt^{\prime}+t^{\prime}\left(  a^{\dagger}+b\right)  +t\left(
a+b^{\dagger}\right)  } &  =\colon e^{t^{\prime}\left(  a^{\dagger}+b\right)
+t\left(  a+b^{\dagger}\right)  }\colon \nonumber \\
&  =\sum_{m,n=0}^{\infty}\frac{t^{m}t^{\prime n}}{m!n!}\colon \left(
a+b^{\dagger}\right)  ^{m}\left(  a^{\dag}+b\right)  ^{n}\colon.\label{b1}%
\end{align}
Comparing with Eq.(\ref{d10}) we see the following identity%
\begin{equation}
H_{m,n}\left(  a+b^{\dagger},a^{\dag}+b\right)  =\colon \left(  a+b^{\dagger
}\right)  ^{m}\left(  a^{\dag}+b\right)  ^{n}\colon.\label{b2}%
\end{equation}
Thus due to $X+Y=\frac{1}{\sqrt{2}}(a+a^{\dag}+b^{\dagger}+b)$, using
Eqs.(\ref{b2}) and we deduce%
\begin{align}
\sqrt{2^{n}}\colon \left(  X+Y\right)  ^{n}\colon &  =\colon \left(
a+b^{\dagger}+a^{\dag}+b\right)  ^{n}\colon \nonumber \\
&  =\sum_{l=0}^{n}\binom{n}{l}\colon \left(  a+b^{\dagger}\right)  ^{l}\left(
a^{\dag}+b\right)  ^{n-l}\colon \nonumber \\
&  =\sum_{l=0}^{n}\binom{n}{l}H_{l,n-l}\left(  a+b^{\dagger},a^{\dag
}+b\right)  .\label{b3}%
\end{align}
From Eq.(\ref{a4}) it is seen that
\begin{equation}
\sqrt{2^{n}}\colon \left(  X+Y\right)  ^{n}\colon=H_{n}\left(  \frac
{a+b^{\dagger}+a^{\dag}+b}{2}\right)  \label{b4}%
\end{equation}
Since $\left[  a+b^{\dagger},a^{\dag}+b\right]  =0$, combining Eq.(\ref{b3})
and Eq.(\ref{b4}) we obtain a binomial-like theorem between the
single-variable Hermite polynomials and the two-variable Hermite polynomials%
\begin{equation}
\sum_{l=0}^{n}\binom{n}{l}H_{l,n-l}\left(  x,y\right)  =H_{n}\left(
\frac{x+y}{2}\right)  .\label{b5-0}%
\end{equation}
In addition, it is seen that
\begin{align}
&  \sum_{m,n=0}^{\infty}\frac{t^{m}t^{\prime n}}{n!m!}\left(  a+b^{\dagger
}\right)  ^{m}\left(  a^{\dag}+b\right)  ^{n}\nonumber \\
&  =\left.  \colon e^{tt^{\prime}+t^{\prime}\left(  b+a^{\dagger}\right)
}e^{t\left(  a+b^{\dagger}\right)  }\colon \right.  \nonumber \\
&  =\sum_{m,n=0}^{\infty}\frac{\left(  -it\right)  ^{m}\left(  -it^{\prime
}\right)  ^{n}}{n!m!}\colon H_{m,n}\left[  i\left(  a+b^{\dagger}\right)
,i\left(  a^{\dag}+b\right)  \right]  \colon,\label{b6}%
\end{align}
this leads to the new operator identity%
\begin{equation}
\left(  a+b^{\dagger}\right)  ^{m}\left(  a^{\dag}+b\right)  ^{n}=\left(
-i\right)  ^{m+n}\colon H_{m,n}\left[  i\left(  a+b^{\dagger}\right)
,i\left(  a^{\dag}+b\right)  \right]  \colon.\label{b8}%
\end{equation}
Based on this, using Eq.(\ref{b5-0}) we have%
\begin{align}
\left(  X+Y\right)  ^{n} &  =2^{-n/2}\sum_{l=0}^{n}\binom{n}{l}\left(
a+b^{\dagger}\right)  ^{l}\left(  a^{\dag}+b\right)  ^{n-l}\nonumber \\
&  =i^{-n}2^{-n/2}\sum_{l=0}^{n}\binom{n}{l}\colon H_{l,n-l}\left[  i\left(
a+b^{\dagger}\right)  ,i\left(  a^{\dag}+b\right)  \right]  \colon \nonumber \\
&  =i^{-n}2^{-n/2}\colon H_{n}\left(  i\frac{a+b^{\dagger}+a^{\dag}+b}%
{2}\right)  \colon \nonumber \\
&  =\left(  i\sqrt{2}\right)  ^{-n}\colon H_{n}\left(  i\frac{X+Y}{\sqrt{2}%
}\right)  \colon,\label{b9}%
\end{align}
which is regarded as the extension of Eq.(\ref{2}). Eq.(\ref{b9}) can also be
proved by using Eqs.(\ref{9}), (\ref{c3}) and (\ref{c4}) as follows%
\begin{align}
&  \left(  X+Y\right)  ^{n}\nonumber \\
&  =\int \frac{d^{2}\xi}{\pi}\left(  X+Y\right)  ^{n}\left \vert \xi
\right \rangle \left \langle \xi \right \vert \nonumber \\
&  =\sqrt{2^{n}}\int \frac{d\xi_{1}d\xi_{2}}{\pi}\xi_{1}^{n}\colon e^{-\left(
\xi_{1}-\frac{X+Y}{\sqrt{2}}\right)  ^{2}-\left(  \xi_{2}-\frac{P_{1}-P_{2}%
}{\sqrt{2}}\right)  ^{2}}\colon \nonumber \\
&  =\sqrt{2^{n}}\int \frac{d\xi_{1}}{\sqrt{\pi}}\xi_{1}^{n}\colon e^{-\left(
\xi_{1}-\frac{X+Y}{\sqrt{2}}\right)  ^{2}}\colon \nonumber \\
&  =\left(  i\sqrt{2}\right)  ^{-n}\colon H_{n}\left(  i\frac{X+Y}{\sqrt{2}%
}\right)  \colon,\label{b7}%
\end{align}
which further proves that the identity in Eq.(\ref{b5-0}) is correct.

In summary, by combining re-ordering operator Hermite polynomials
and the IWOP technique we can we directly derive new identities and
new integration formulas without really doing the integration in the
usual sense, this is useful for developing Newton-Leibniz
integration performed on Dirac's ket-bra operators.

\end{document}